\documentclass{article}

\usepackage{PRIMEarxiv}

\usepackage[utf8]{inputenc} 
\usepackage[T1]{fontenc}    
\usepackage{hyperref}       
\usepackage{url}            
\usepackage{booktabs}       
\usepackage{amsfonts}       
\usepackage{nicefrac}       
\usepackage{microtype}      
\usepackage{lipsum}
\usepackage{fancyhdr}       
\usepackage{graphicx}       
\graphicspath{{media/}}     

\title{A Machine Learning Approach to Automatic Classification of Eight Sleep Disorders
}

\author{
  Dylan Zhuang and  Ivey Rao \\
  Research Intern, EECS Department\\
 Florida Atlantic University\\
 Boca Raton, FL, USA\\
  \texttt {dzhuang2018@fau.edu, ivey.rao@pinecrest.edu} \\
   \And
 Ali K Ibrahim \\
 EECS Department and HBOI\\
 Florida Atlantic University\\
 Boca Raton, FL, USA\\
  \texttt{aibrahim2014@fau.edu} \\
}

\begin{document}
\maketitle

\begin{abstract}
This research focuses on automatically classifying common sleep disorders. We attempt to answer the following basic research questions: Is a machine learning model able to classify all types of sleep disorders with high accuracy? Among the different modalities of sleep disorder signals, are some more important than others? Do raw signals improve the performance of a deep learning model when they are used as inputs? 
Prior research showed that most sleep disorders belong to eight categories (for instance, the PhysioBank dataset). To study the performance of machine learning models in classifying polysomnography recordings into the eight categories of sleep pathologies, we selected the Cyclic Alternating Pattern (CAP) Sleep Database, a collection of 108 polysomnographic recordings. We developed a multi-channel Deep Learning (DL) model where a set of Convolutional Neural Networks (CNNs) were applied to six channels of raw signals of different modalities, including three channels of EEG (lectroencephalogram) signals and one channel each of EMG (electromyography), ECG (electrocardiogram), and EOG (electrooculogram) signals. To compare the performance of the DL model with other models, we designed a model that took spectral features, instead of raw signals, as its inputs. We named the former the DL-R model and the latter the DL-F model. For comparison, we also designed two conventional classifiers, Random Forest (RF) and Support Vector Machines (SVM), which also took spectral features as inputs.
We first studied the "importance" issue of signal modalities using the RF algorithm. We found that ECG contributed most to the important features and EMG second, among the four signal modalities. We then studied the accuracy performance of the proposed machine learning models. We verified that the multi-channel DL-R model, which took raw signals as its inputs, outperformed all other models, with its sensitivity and specificity scores both being above 95\% across all the eight sleep disorders. This accuracy performance is on a par with those published results which dealt with fewer types of sleep disorders. To study the explanability of the DL-R model, we adopted two popular heatmap-generating techniques, with which we confirmed that the DL model's superior performance was owing to the CNN network's ability to extract potent features from raw signals. Its heatmaps produced frequency ranges and peak frequencies of various sleep disorders consistent with the results from clinic studies. We hope that the proposed approach is one step closer to more able and trustworthy machine learning techniques that one day will be adopted by practitioners.
\end{abstract}


\section{Introduction}

\noindent Many sleep disorders lead to cardiovascular, metabolic, and psychiatric issues, affecting over 100 million Americans \cite{A7, 1}. Narcolepsy, sleep-disordered breathing, insomnia, periodic leg movements (SDM), bruxism, rapid eye movement behavior disorder (RBM), and nocturnal frontal lobe epilepsy are among these disorders (NFLE) \cite{1}. Several techniques have been widely used to monitor patients' sleep patterns during a polysomnography test. For example, the electroencephalogram (EEG) records brain wave activity, the electrooculogram (EOG) records eye movement, the electromyography (EMG) records muscle movement, and the electrocardiogram (ECG) records the heart's electrical activity. 
Due to time constraints  Visiting a sleep clinic for a polysomnography (PSG) test is not always feasible. Professionals must manually examine sleep recordings, a time-consuming task \cite{11}. In \cite{28}, PSG recordings were scored according to R and K rules to classify sleep stages, where less than thirty-second epochs were recommended by the R and K guidelines. For optimal alpha and spindle viewing, the 30-second interval was historically used, as reading one page takes 30 seconds at ten mm/s. Each epoch was given a stage, and whenever two or more stages overlapped, the stage with the most time was scored. A literature review from 2010-to 2020 found that using other PSG recordings besides EEG could aid in more accurate diagnoses. A non-rapid eye movement sleep event that has been associated with several disorders and is a defining feature of sleep instability that may be detected with an EEG. However, this approach generates a vast amount of data during a full night test, making manually scoring all the cyclic alternating pattern cycles impractical and prone to error \cite{29}. In addition, the high-dimensional data of PSG recordings may cause overfitting in programmed diagnostic tools for sleep disorders, especially those based on machine learning. Instead of using raw data directly to diagnose sleep disorders, researchers applied feature extraction techniques to feed classification models (i.e., neural networks). However, when data is reduced in size, essential information for sleep disorder diagnoses may be lost. A literature review also revealed certain limitations of the existing machine learning models in classifying sleep disorders using PSG recordings.  
Many notable studies regarding classifying sleep signals using Deep Learning (DL), a type of machine learning, have been reported. An early attempt at using such an approach for image classification can be found in \cite{A1}. One model had an attention mechanism built-in, which improved the feature extraction performance of the model through intra and inter-epoch feature learning, which reached an accuracy of 82.8\% \cite{15}. Another model used fine-grained methods, constructing time series from EEG epochs and then stringing together related ones to form time series. It reached an accuracy of 92.2\% \cite{16}. Nevertheless, another study used a 19-layer 1D Convolutional Neural Network (CNN) with ten convolutional layers, which reached a 90.8\% accuracy \cite{17}. More recent studies using DL models, including CNNs, for analyzing and classifying sleep disorders can be found in \cite{urtnasan2021ai, A2, A3, A4, exarchos2020supervised}. Notably, \cite{urtnasan2021ai} achieved remarkable results with a single CNN model for the automatic classification of four sleep disorders.\\For sleep disorder classification, Recurrent Neural Networks (RNNs) and Long Short-Term Memory networks (LSTMs) were less commonly reported in the literature. However, an end-to-end hierarchical LSTM with an attention-based recurrent model with filter bank layers reached an accuracy of 87.1\%, higher than other LSTM or RNN models \cite{18}. Hybrid models, which combined CNNs with either RNN or LSTM architectures, were also reported. In such models, convolutional layers were employed at the front to extract features, and RNN or LSTM layers at the back to recognize patterns in feature maps \cite{19, 20}.   
\\For future directions of research, we reccomend signals from different instruments, EEG, EOG, and EMG, for predicting sleep phases with DL models, \cite{14}. In the literature, numerous studies chose CNNs and used EEG signals as inputs due to CNNs' effectiveness at recognizing characteristic features of sleep EEG signals. We used 1D CNNs more often than 2D and 3D CNNs, because the input signals are one-dimensional time series. 
A few studies used other PSG signals alongside EEG signals, but these studies did not perform as well as those that used only EEG signals. Cyclic Alternating Pattern (CAP) detection should be focused on, as EEG signals mark irregular sleeping patterns \cite{14}. However, it was not easy to compare the performance of different models and approaches because most were trained on only one database. In addition, most of the published results focused only on a limited number of sleep disorders. 
\\In this paper, we explore the potential applications of machine learning techniques, emphasizing deep learning techniques in their efficacy to identify a wide range of sleep disorders using multiple modalities of signals. First, we will show that models constructed with DL algorithms can be trained directly with raw signals of PSG recordings. In other words, feature extraction is not required in this case. Because raw data is used to train the classification model, information loss is no longer a significant problem. We will show that DL models can outperform traditional machine learning algorithms such as Support Vector Machines (SVMs) and Random Forests (RFs)  in one of the most critical measures, identifying actual patients of various sleep disorders. We also studied the issue of feature importance related to multiple modalities of PSG measurements. Finally, we addressed the explainability of the proposed DL model with heatmaps generated by two popular techniques.
The remainder of this paper is organized as follows. Section 2 reviews the essential technical tools used in this study. Section 3 details the design of the multi-channel deep learning model, DL-R, among other things. Section 4 presents experimental results with discussions, including those from the feature importance and model performance comparison studies. Concluding remarks are given in the last section.

\section{Preliminaries}
\noindent This section reviews essential technical tools relevant to this study, including spectral analysis, basics of machine learning and deep learning, and techniques for creating heatmaps for the explainability of deep learning models.  

\subsection{Spectral Analysis}
\begin{figure}[p]
    \centering
    \includegraphics[scale=.16]{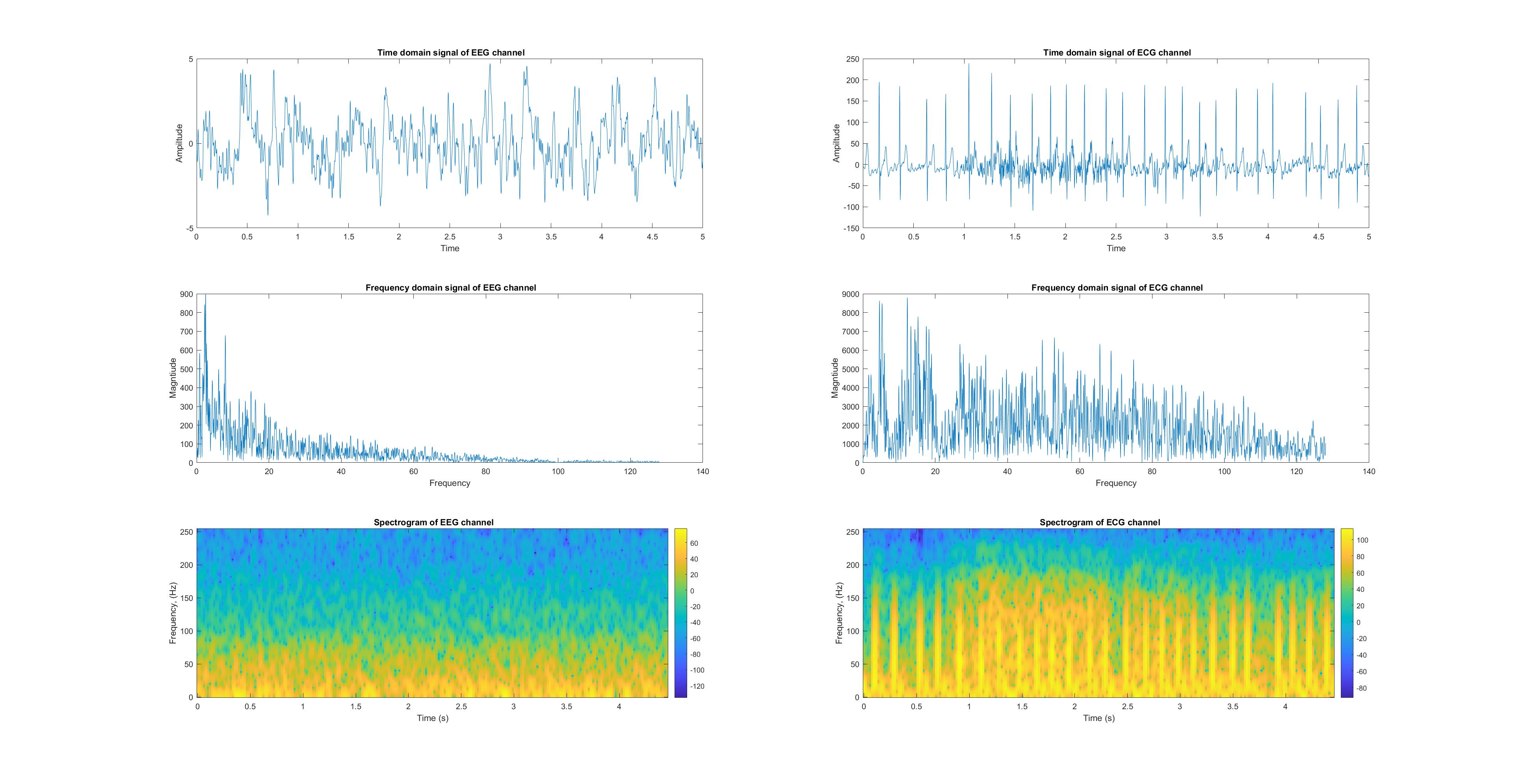}
    \caption{Time signals (first row), their frequency counterparts (middle row), and their spectrograms (bottom row) of EEGs and ECGs. }
    \label{sig1}
\end{figure}

 \noindent We analyzed power spectra of the EEG, ECG, EOG, and EMG of all sleep disorders in our dataset. Additionally, time-frequency representations were used in the analysis. Signals are converted to spectrograms using Short Time Fourier Transform (STFT), where Fourier Transform converts a signal from a time domain to a frequency domain. Spectrograms are created by overlapping 0.2s Hanning windows, or 101 samples at 512 samples per second for these signals. Furthermore, each frame is subjected to an 8192-point Fast Fourier Transform (FFT) with a 0-80 Hz frequency range. Figure \ref{sig1} illustrates time signals, their frequency representations, and their spectrograms of EEG and ECG measurements. Similar plots can be obtained for signals of other modalities. From our analysis, we found that the peak frequency of bruxism (Bru) ranges from 20-30 Hz, insomnia (Ins) peaks from 10 Hz to 20 Hz, narcolepsy (Nar) spikes around 10 Hz, nocturnal frontal lobe epilepsy (Nfl) spikes at 10 Hz, periodic leg movements (PLM)  peaks around 5 Hz, REM behavior disorder (Rbd) peaks around 15Hz, and sleep-disordered breathing (Sdb) peaks around 40 Hz; refer to Table \ref{T1}.

\begin{table}[]
 \footnotesize
\caption{\label{T1}Frequency Range and Peak Frequency Range of Various Sleep Disorders}

\begin{center}\begin{tabular}{c c c  }
\hline\hline 
 Sleep Disorder  & Frequency Range (Hz) & Peak Frequency Range (Hz) \\
  \hline
Bru & 0-50 & 10-25\\
Ins & 0-100 & 5-45\\
Nar & 0-50 & 8-13\\
Nfl & 0-50 & 8-15\\
Plm & 0-50 & 2-7\\
Rbd & 0-50 & 10-20\\
Sbd & 0-50 & 30-45\\

\hline\hline
\end{tabular}
\end{center}
\label{T:1}
\end{table}

\subsection{Machine Learning}
\noindent The field of machine learning is rapidly-advancing, as both techniques and applications are constantly being discovered and improved. This section attempts to outline basic methods in machine learning relevant to this study, except deep learning, which will be discussed in a follow-up section.

\subsubsection{Support Vector Machine}
\noindent The goal of support vector machines, a supervised machine learning algorithm, is to find an optimal hyperplane or decision boundary which clusters data and classifies data points effectively through optimizing the distance between the hyperplane and those feature points of the dataset which are nearest to the decision boundary (support vectors). SVMs can rapidly classify data using hundreds or thousands of features \cite{24}.
Support Vector Machines are a simple, intuitive, and robust method for data classification and regression analysis, as they may be applied to both linear and non-linear classification. In the latter case, the kernel trick is employed for effective pattern analysis. SVMs are often used to obtain a baseline performance for machine learning-based models. Readers are referred to  \cite{24} for more details on SVMs.
\subsubsection{Random Forest}
\noindent Random forest, a supervised machine learning method, employs a large number of relatively uncorrelated, small trees in the training phase. When a trained random forest is used for classification, it selects the class to which the data point belongs with the collective opinions of these trees. For regression tasks, each tree's prediction contributes to the overall prediction of the random forest. A random choice forest is utilized to counteract a decision trees' propensity for overfitting to its training set. The trees are created using various strategies, including bootstrap aggregation or bagging, a method for minimizing the variance of a prediction function in which random points from the dataset are picked without replacement and used to construct decision trees. Because the samples utilized in the bagging process come from the same dataset as the data used to train the trees, the input and target data are identically distributed. The purpose of bagging is to generate a large number of varied but unbiased trees and then average their outputs or follow the categorization that most trees voted. Surprisingly, when trees are fitted with additional layers, their variability dramatically rises while being less biased. As a result, random forests may outperform their trees in most cases. However, random forests can still suffer from the overfitting syndrome, which becomes more likely as the size of the random forest expands. Another type of tree selection and growth is "boosting," Trees are developed adaptively without identical distribution. The method provides a great potential for performance improvement since minimizing variability is a feasible approach for improving bagging distributions. As a result, as the tree count grows, boosting starts to outperform bagging \cite{30}.

\subsection{Deep Learning}
\noindent Deep learning is a subset of machine learning that implements a deep neural network with a large layer count to extract more complex features from input signals. While an external neural network contains many connected neurons with a couple of layers, a deep neural network can be composed of dozens to hundreds of layers with millions of neurons. 
\subsubsection{Convolutional Neural Network}
\noindent A type of deep neural network used in our research is Convolutional Neural Network, a commonly accepted image recognition and classification model. \cite{gu2018recent} CNNs became popular after the exceptional performance of AlexNet in the 2012 ImageNet competition \cite{A1} and have been proven to be one of the most effective deep learning algorithms for various applications. \cite{gu2018recent} A CNN consists of these major building blocks: convolution, pooling, activation, and batch normalization. An input signal is initially fed into a convolution layer in the form of a tensor, a group of convolutional kernels with variable dimensions and manually adjustable hyperparameters such as step and padding. A convolution layer involves a process similar to the neuron's function in one's visual cortex, as it convolves inputs and sends them to the next layer. After the input passes through a convolution layer, activation, pooling, and batch normalization functions are applied to simplify the convolutional outputs. This process is repeated several times, depending on the depth of the neural network. Finally, pooling and fully connected layers flatten the network outputs to provide a single vector of features for classification. Due to the nature of complex time series, there may be little distinction in input signals which may be hugely varied in their underlying frequencies, which is why we use spectrograms as inputs to the CNNs to create more distinct representations of auditory data.

\subsection{Grad-CAM}
\noindent To improve the explainability and transparency of the classification model, we employed a Matlab module called "Grad-CAM," which attempts to produce visualizations of the decisions made from the archetypical Convolutional Neural Network architecture. In science, computing, and engineering, the phrase "black box" relates to a model's level of procedural disclosure; in this instance, it exposes nothing. Businesses use the "black box" technology to safeguard intellectual property and preserve competitiveness. When a system cannot explain how it arrived at an answer, this is referred to as "the black-box problem" \cite{31}. The "Grad-Cam" technique proposes a procedure for creating "visual explanations" for choices made by a large class of CNN-based models, so making them more transparent \cite{27}.
\\\\
\noindent Grad-CAM is a post-hoc explanation approach that may be used to any CNN without modifying the network under examination \cite{27}. Grad-CAM creates a heat map of class activation when given an input picture and a class: it color codes regions of the input image according to their positive contribution to the system identifying the image with the specified class, outputting a class activation map. In this study, the activity intensity ranges from orange (most active) to blue (least active).
\subsection{LIME}
\noindent Local interpretable model-agnostic explanations (LIME) is an approach for providing explanations for any classifier. It provides an easily interpretable model that is locally close to the true model for a given input and forecast \cite{ribeiro2016should}. By highlighting superpixels that contribute positively to the classification in orange and negatively in blue and providing a quantifiable measure of the contribution, the LIME technique can segment an image and demonstrate which portions of it contribute the most positively to the classification in an attempt to make light of a "black box."

\section{Dataset, Feature Extraction, and DL Model Design}
\noindent In this section, we first introduce the dataset used in our experimental study. After this, we discuss the feature extraction procedure for the RF and SVM classifiers. We then present the architecture of the DL classification model. 
\subsection{Dataset}
\noindent The dataset used in the experimental study, PhysioBank, is a databank created by PhysioNet to encourage exploratory research in treating cardiological diseases using technology and facilitating relevant clinical studies. It provides a comprehensive database that was previously not readily available, and it also provides data to nonspecialists to aid research that may lead to the discovery of novel methods through interdisciplinary efforts. The dataset was initially created in 1999 with a grant from the National Institute of Health and has been consistently updated since then. It is a freely accessible dataset containing many clinical and biological data. 
\\
Our research utilizes the multi-parameter waveform database with three channels of EEGs, one channel each of EOG, EMG, and ECG, and data collected from patients with Bruxism, Insomnia, Narcolepsy, Nocturnal Frontal Lobe Epilepsy, Rapid Eye Movement Behavior Disorder, and sleep-disordered breathing. Expert Neurologists provided scoring of patient sleep macrostructure, while data such as sleep stage, body position, time of day, and sleep duration were documented in a separate text file \cite{25}. With the rapid advancement of deep learning algorithms and computational capability, it is now possible for us to process such a large dataset with many enumerated and complex parameters.

\subsection{Feature Extraction for RF and SVM Algorithms}
\noindent For a comparison study, the two conventional machine learning algorithms, the SVM and RF algorithms, are used in this research. For these two algorithms, spectral features of input signals were extracted. For this purpose, we converted each channel into frequency components using an FFT procedure with NFFT = 8192 resolution. Each spectrum was divided into five non-overlapped bands ranging from 0 Hz to 50 Hz since, from spectral analysis, most information was concentrated in this frequency range. The width of each band was 10 Hz. Nine spectral features were extracted from each band. These features include power, mean, standard deviation, median, minimum, maximum, entropy, and energy calculated from the spectrum of each band. Since there were six channels of signals, the total number of features was then $5 \times nine \times 6 = 270$ for 5 seconds. 

\subsection{Design of the DL Classification Model}
\begin{figure}[p]
    \centering
    \includegraphics[scale=.4]{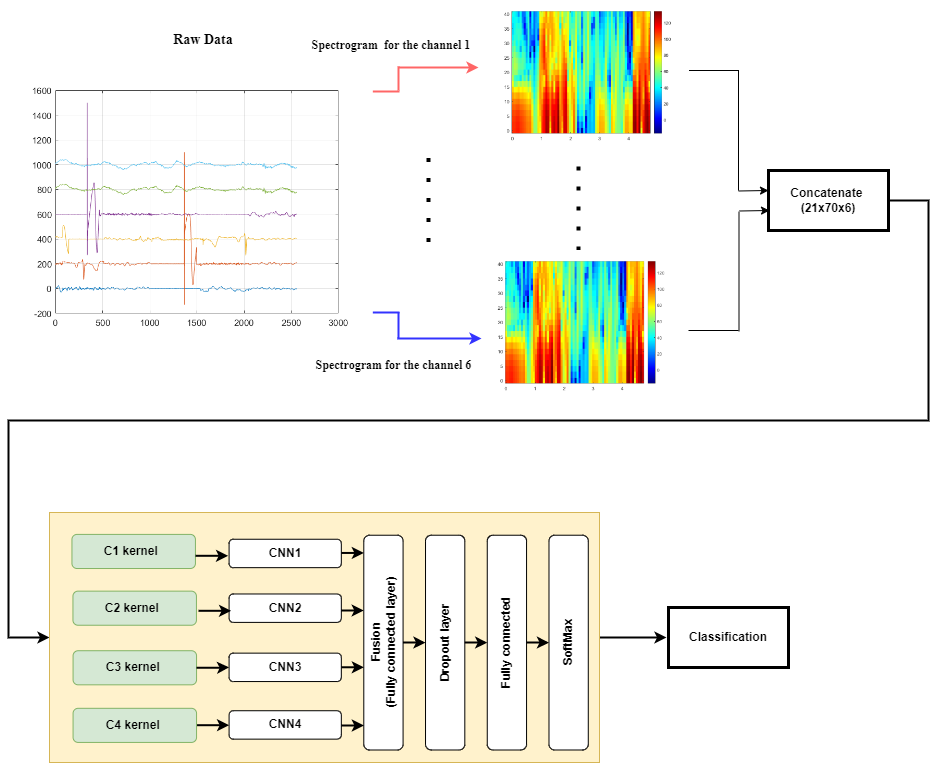}
    \caption{Sleep disorder classification model with raw data as inputs (DL-AR). It performs mainly the following tasks: converting the signals into spectrograms, extracting discriminative patterns from individual spectrograms using CNNs, and classifying 5s frames into different classes using a softmax classifier. The input signals include three channels of EEG, one channel each of EMG, EOG, and ECG signals. The three EEG signals are fed to one CNN model, and the EMG, EOG, and ECG are each fed to one of the remaining CNN models.}
    \label{sys}
\end{figure}

\noindent A description of the DL proposed model for sleep disorder classification is given in this section. Since the DL classification model takes raw inputs directly, it is denoted as the DL-AR model. The system diagram shown in Figure \ref{sys} consists of mainly the following functional blocks: time-frequency representation,  CNNs, and fusion mechanism. 
\\\\
In the proposed method, raw signals from three EEG channels, one EMG channel, one EOG channel, and one ECG channel, are first converted to spectrograms using STFT. As mentioned, a spectrogram is an image of a signal's time-varying spectral content over a range of frequencies. This work generates spectrograms by framing the signal with 0.2 s Hanning windows, or 101 samples at a sample rate of 512 samples per sec and 50\% overlap. Then, an 8192-point Fast Fourier transform (FFT) is applied to each frame. The spectrogram is estimated by taking the magnitude of each FFT frame with a frequency range of 0–80 Hz.
\\\\
 After the signals are converted into six channels spectrograms with the size of 21x70 each, We designed four different convolutional layers that separate the six channels into four outputs. These outputs correspond to EEGs, EMG, ECG, and EOG channels. The first three channels were fed into a single CNN, while the remaining channels were input into separate CNNs. Each CNN subnet extracts features from input channels that contain comprehensive information related to sleep disorders. The outputs of these subnets are connected to a fully connected layer and subsequently fused with a dropout layer and a fully connected layer. Finally, a SoftMax layer classifies each 5s frame signal segment into a standard frame or a frame corresponding to a particular sleep disorder. The CNNs in the model are identical internally, except that CNN1 has three channels, and the rest has only one each. The parameters of the CNN models are given in Table \ref{T2}.  
 \\\\
For comparison, a DL model taking the features (spectra) produced by the Fourier Transform as inputs was also designed (DL-F). The model had an identical structure to the DL-F model depicted in the bottom part of Figure \ref{sys}, except that in this case, four one-dimensional instead of four two-dimensional CNNs were used in the model with different hyperparameters. This is because the features extracted by the Fourier Transform are one-dimensional. The parameters of the DL-F model are given in Table \ref{T3}. 
 
\begin{table}[p]
 \footnotesize
\caption{\label{T2}Hyperparameters of CNN1 in the DL-R Model}
\begin{center}\begin{tabular}{c c c  c}
\hline\hline 
 Layer Number  & \#  Conv Filters & Kernel Size & Output Dimension\\
  \hline
  Input Layer & $21\times 70\times 3 $ &---&---\\
Cov1+ReLu1+Maxpooling1 ($2\times 2$ w. stride [2,2]) & 32 & $3\times 3$ & $10\times 35 \times32$\\
Cov2+ReLu2+Maxpooling2 ($2\times 2$ w. stride [2,2])& 64 & $3\times 3$& $10\times 17 \times64$\\
Cov3+ReLu3+Maxpooling3 ($2\times 2$ w. stride [2,2]) & 128 & $3\times 3$ & $10\times 9 \times128 $\\
Cov4+ReLu4+Maxpooling4 ($2\times 2$ w. stride [2,2]) & 256 & $3\times 3$ &$ 5\times 5 \times256$\\
Cov5+ReLu5+Maxpooling5 ($2\times 2$ w. stride [2,2]) & 512 & $3\times 3$ & $3\times 3 \times512$\\

\hline\hline
\end{tabular}
\end{center}
\label{T:1}
\end{table}
 
 \begin{table}[p]
 \footnotesize
\caption{\label{T3}Hyperparameters of CNN1 in the DL-F Model}
\begin{center}\begin{tabular}{c c c  c}
\hline\hline 
 Layer Number  & \# Conv Filters & Kernel Size & Output Dimension\\
  \hline
Input Layer & $2000\times 1\times 3$ &---&---\\
Cov1+ReLu1+Maxpooling1 (4,1 w. stride 4) & 32 & $5\times 1$ & $500\times 1 \times 32$\\
Cov2+ReLu2+Maxpooling2 (4,1 w. stride 4) & 64 & $5\times 1$ & $125\times 1 \times64$\\
Cov3+ReLu3+Maxpooling3 (4,1 w. stride 4)& 128 & $5\times1$ & $32\times 1 \times128$\\
Cov4+ReLu4+Maxpooling4 (2,1 w. stride 2) & 256 & $3\times 1$  &$ 16\times 1 \times256 $\\
Cov5+ReLu5+Maxpooling5 (2,1 w. stride 2) & 512 & $3\times 1$ & $8\times 1 \times512$\\

\hline\hline
\end{tabular}
\end{center}
\label{T:1}
\end{table}

\section{Results and Discussions}
\noindent In this section, we first discuss the model testing procedure and the metrics used to evaluate various models' performance. We then demonstrate the feature importance using the RF model and look into the DL classification model's explainability with heatmaps generated with two popular techniques. Finally, we show the performance of the classification models with discussion.

\subsection {Model Testing Procedure and Evaluation Metrics}

\noindent The experiments were implemented in Matlab. First, a standard validation procedure was followed \cite {FDA, A6}. Next, the experimental dataset from PhysioNet was divided into 80\% and 20\%, where 80\% of the data was used for training, and 20\% of the data was reserved for testing. Among the training data, 80\% of the data were randomly chosen for training and the remaining 20\% for validation. This step was repeated five times until all data points in the training set were validated once. The trained model was then tested with the reserved test data.
\\\\In this study, we use sensitivity and specificity to measure the performance of various classification models. The ability of a model to identify true positives is referred to as sensitivity, while the ability of a model to identify the null is referred to as specificity. More specifically, the sensitivity score would show the ability of the model to correctly identify the candidate's disease, while the specificity score would show the ability of the model to state that the candidate does not have a disease correctly. 

\subsection {Feature Importance and Model Explanability}

\begin{figure}[p]
    \centering
    \includegraphics[width=8 cm]{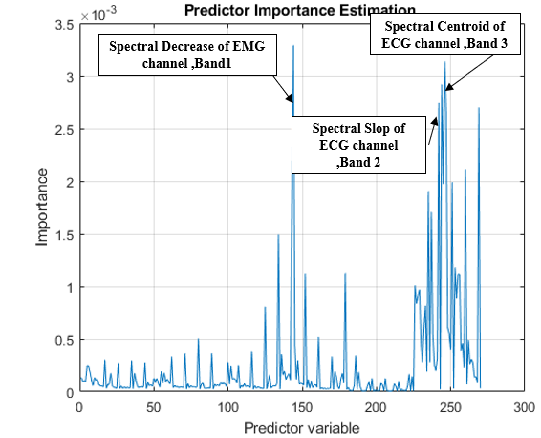}
    \caption{Results of the feature importance analysis with the RF algorithm. It can be seen that the more important features are around number 250, corresponding to the ECM region, and the number of important features around 150, corresponding to the EMG region, ranks the second.}
    \label{fig:importance}
\end{figure}

\noindent To understand which features played more significant roles in sleep disorder classification, features importance measures were calculated using the RF algorithm 
 by assessing the node risk (i.e., change in node impurity weighted by the node probability) associated with
splitting the data using each feature \cite{RF}. The weights of the most important features were shown in Figure  \ref{fig:importance} and Table \ref{T:5}, the latter of which lists the 16 most essential features ranked by the RF algorithm. It can be seen that the most critical features of the RF model are the spectral decrease of the EMG channel and the spectral centroid of the ECG channel (Figure \ref{fig:importance}). Furthermore, The ECG modality provides more important features than other modalities (Table \ref{T:5}), followed by the EMG modality.

 \begin{table}[p]
 \footnotesize
\caption{\label{T:5} Important Features Identified by Random Forest}
\begin{center}\begin{tabular}{c c c}
\hline\hline 
 Feature  & Band & Channel\\
  \hline
Spectral Slope & 5 & EEG\\
Spectral Slope & 1 & EMG\\
Spectral Dcrease & 1 & EMG\\
Spectral Slope & 2 & EMG\\
Spectral Slope & 5 & EMG\\
Spectral Centroid & 1  & ECG\\
Spectral Centroid & 2  & ECG\\
Spectral Skewness & 2  & ECG\\
Spectral Slope & 2  & ECG\\
Spectral Centroid & 3  & ECG\\
Spectral Skewness & 3  & ECG\\
Spectral kur & 3  & ECG\\
Spectral Entropy & 3  & ECG\\
Spectral Crest & 3  & ECG\\
Spectral entropy & 4  & ECG\\
Spectral Centroid & 5  & ECG\\
\hline\hline
\end{tabular}
\end{center}
\label{T:5}
\end{table}

\begin{figure}[p]
    \centering
    \includegraphics[width=14cm]{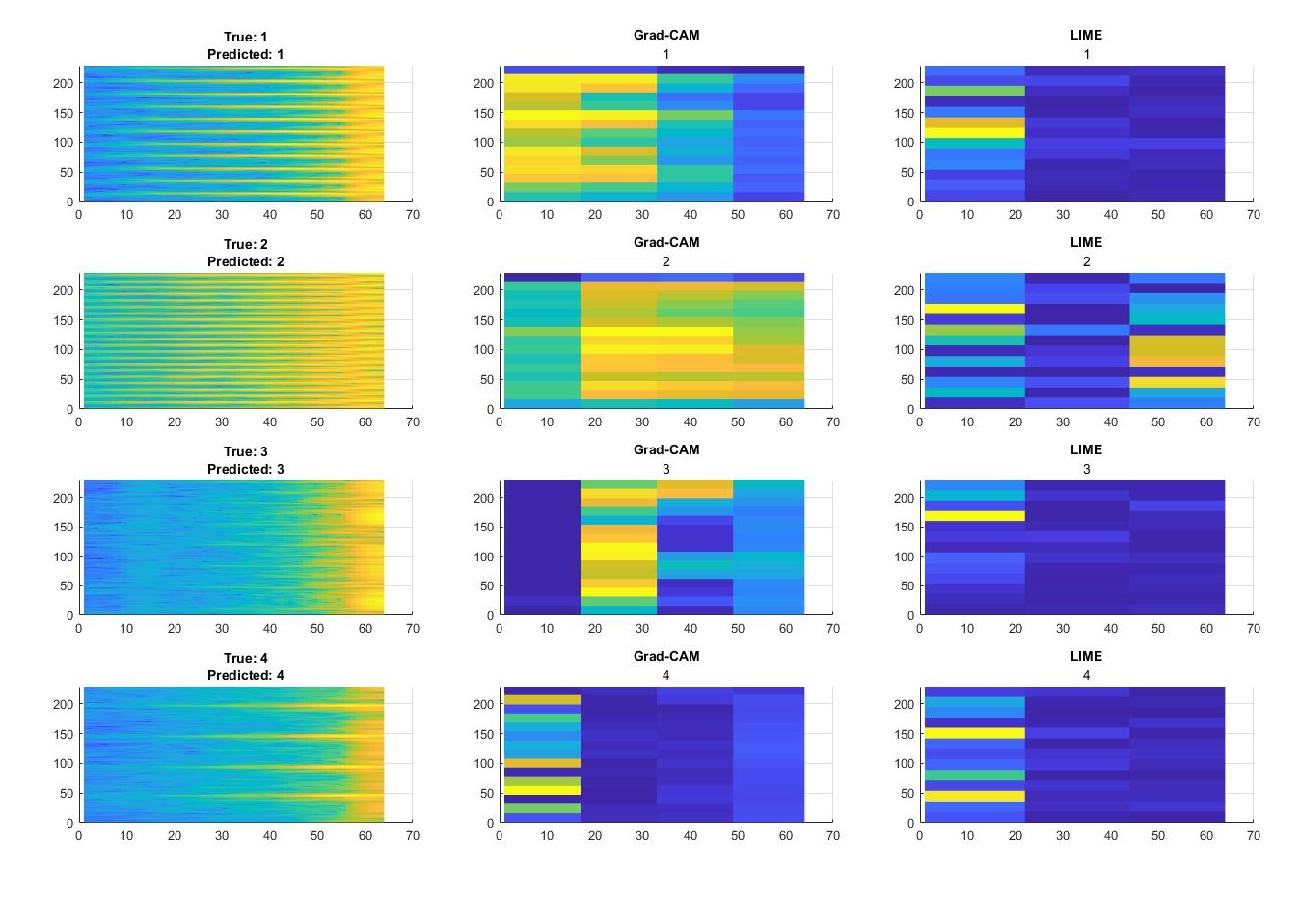}
    \caption{The heatmaps created by of Grad-Cam, and LIME Techniques. Row 1 to row 4 illustrate spectrograms of different disorders (Bru, Ins, Nar, and Nfl) of ECG signals and their corresponding heatmaps generated from the spectrograms. Note that the horizontal axis shows the frequency in Hz, and the vertical axis shows time in seconds.}
    \label{fig:heatmap}
\end{figure}

\noindent To show the discriminative ability of the proposed DL model, heatmaps using Grad-CAM and LIME Techniques were also created; refer to  Figure \ref{fig:heatmap}. We only show the results of four sleep disorders: Bru, Ins, Nar, and Nfl. The plots corresponding to the remaining disorders were omitted here because these exhibited similar behavior. Examining the figure, one can observe that different sleep disorders exhibit distinct patterns. For instance, concerning the Grad-CAM technique, Bru and Nfl have the most active regions in the low-frequency range, with Bru showing more activity.
On the other hand, disorder Ins is more active in a wide frequency range and Nar in the middle-frequency range. These results confirm the data shown in Table \ref{T1}. LIME maps are given in the figure and illustrate the CNN model's capabilityodel in its discriminative ability. 

\subsection {Sensitivity and Specificity of the Classification Models}

\noindent The classification model outlined in Figure \ref{sys}, denoted as DL-AR in Tables \ref{T:2} and \ref{T:3}, was tested extensively on the PhysioNet dataset. For comparison, we also used Fast Fourier Transform (FFT) as another way to extract features from sensory signals. The resulting model was denoted as DL-F. Sensitivity is the most important metric since it measures the ability of a model to classify true disorders. We computed sensitivity scores for various models first, shown in Table \ref{T:2}. With the DL-F model, which used an FFT algorithm for feature extraction for the DL algorithm, we achieved a sensitivity score ranging from 66.1\% for patients with periodic leg movements to 98.1\% for patients with bruxism. On the other hand, with the DL-R model, which used spectrograms as inputs to DL algorithms, we achieved a sensitivity score ranging from 94.8\% for patients with REM behavior disorder to 99.6 for patients with bruxism.In summary, the DL-R model achieved over 95\% accuracy in identifying all nine sleeping patterns (eight disorders and a normal use case). 
\\To show the effectiveness of the proposed DL model, we also compared the results with two conventional machine learning algorithms, SVM and RF. Table \ref{T:2} shows that the DL-R model also outperformed the SVM and RF models by a large margin for identifying patients with various sleeping disorders. More strikingly, the DL-R model performs better in identifying patients with periodic leg movements. In addition, the performance of DL-R exceeds that of any other literature observed \cite{16,17,18}, including various RNN and LSTM models.
\\While specificity, which shows a model's ability to identify patients who truly do not have sleeping disorders, is not as important as the sensitivity measure, it is still useful. Therefore, we computed the specificity scores of the models above, and the results are listed in Table \ref{T:3}. It can be observed that the DL-R, SVM, and RF models all achieved high scores. On the other hand, the performance of the DL-F model is not nearly as good. 
\\We hypothesize that the superiority of the DL-R model is due to the ability of a DL algorithm in handling a vast amount of data. When a set of hand-crafted features are fed to a DL model, like in the case of the DL-F model, it has particularly low specificity and sensitivity for patients with periodic leg movements. It must be noted that despite the relatively low metrics of the DL-F model, i.e., a low sensitivity for patients with periodic leg movements and low specificity for normal patients, but a high sensitivity for other sleeping disorders, suggesting that the model produced a high true positive score but also many false positives. The DL model, in which the CNN networks extract features from raw signals, drastically reduced the number of false positives. 


\begin{table}[p]
 \footnotesize
\caption{\label{T:2}Comparative Sensitivity of Different Models}

\begin{center}\begin{tabular}{cccccc }
\hline\hline 

Disorder& SVM (\%) &RF (\%)&DL-F (\%){0.15 in} &DL-R (\%) &\\
  \hline
Bru & 90.4 & 94.9 & 98.1 &{99.6}\\
Ins&79.4&80.6& 85.6& {98.1} \\

Nar&80.3&88.4&84.3 & {97.6} \\

Plm&62.9&67.4& 66.1 &{96.0} \\

Rbd&86.6&90.0 & 82.4 &{94.8}\\

Sbd&95.4&96.3& {96.8} &96.1 \\

Nfl&77.9&85.8 & 96.4 & {97.0} \\

Nrm&79.9&82.8 & 89.0& {98.2}\\

\hline\hline
\end{tabular}
\end{center}
\label{T:3}
\end{table}
\begin{table}[p]
 \footnotesize
\caption{\label{T:3}Comparative Specificity of Different Models}
\begin{center}\begin{tabular}{c c c c c }
\hline\hline 

 Disorder& SVM (\%){0.15 in} &RF (\%)& DL-F (\%){0.15 in} & DL-R (\%) \\
  \hline
Bru&99.7&99.8& 96.9&{99.6} \\
Ins&98.7&{99.0}& 82.5&95.9\\
Nar&95.3&94.7& 75.7&{96.2} \\
Plm&96.5&{98.0}& 87.1&95.6\\
Rbd&96.7&98.1 &97.7& {98.4}\\
Sbd&98.0&97.5& 89.4&{99.9}\\
Nfl &97.2&{98.1}& 62.1&95.7 \\
Nrm &97.1&{98.7}& 95.9&95.9\\

\hline\hline
\end{tabular}
\end{center}
\label{T:1}
\end{table}

\section{Conclusions}

\noindent Sleep disorders affect over one-third of the general American population, with numerous physical and mental adverse effects. Due to time limitations, visiting a sleep clinic for a polysomnography (PSG) test is not always practical. Professionals must check sleep records manually, which is a time-consuming and financially strenuous procedure. Automatic classification of sleep disorders using machine learning techniques has been seen as a potential solution. By testing various classical machine learning and deep learning techniques in this study, we provided a potential framework for attributing sleep disorders to PSG outputs, using a novel multi-channel Convolutional Neural Network architecture to classify raw data. This study established that deep learning can accurately process multi-channel biometric data, in this case, brain and body activities occurring during non-rapid eye movement sleep. 
\\Observing the Random Forest techniques' results, we noticed that the channels with the most identified features were the ECG and EMG ones, suggesting that among the multi-channel data, the most discriminative signals originated from the heart, muscle, and nerve data. In addition, heatmaps generated from the deep learning classification model agreed with the clinical observation in terms of the frequency ranges and peak frequencies of various sleep disorders. For instance, We observed that different sleep disorders exhibit distinct patterns in the heatmaps. Bru and Nfl are in the low-frequency range and the former displays more intense activity. Furthermore, Ins is active in the middle and up frequency ranges, and Nar is only active in the middle-frequency range.       
\\With both sensitivity and specificity scores above 95\% for all the sleep disorders included in the study, the deep learning model with raw data as inputs is the best-performing technique among all the models tested. Those feature-based techniques, including Random Forest, Support Vector Machines, and the deep learning model with spectral features as inputs, performed poorly, especially sensitivity. This is in agreement with the observation given by other researchers \cite{sainath2015learning,dai2017very}. The results observed have demonstrated great promise in the application of deep learning in classifying medical data, in comparison to the results reported in the literature \cite{urtnasan2021ai,exarchos2020supervised,SALARI2022115950,bandyopadhyay2022clinical}. We  compared more closely our findings to those in \cite{urtnasan2021ai}. The latter reflected the literature's frontier. The difference is that \cite{urtnasan2021ai} examined four sleep disorders: Ins, PLM, Rbd, and Nlfe.
On the other hand, our research covered all (8) sleep disorders in the PhysioBank dataset. We added Bru, Nar, Sbd, and Nfl to the list of those analyzed by \cite{urtnasan2021ai}. These last four sleep disorders are also clinically significant, and hence their inclusion in this and future investigations is justified \cite{A7}. 
\\ The idea of using multi-channel deep learning classifiers with raw data has potential for a wide range of medical applications, yet the scope of this paper is limited as it only studies a narrow set of medical data. Further research should seek to develop deep learning techniques which can deal with varying types of signals while also focusing on establishing more accurate and more trustworthy models to be acceptable to practitioners.

\bibliographystyle{unsrt}  
\bibliography{references}

\begin{thebibliography}{10}

\bibitem{A7}
Thorpy~M. J.
\newblock Classification of sleep disorders. neurotherapeutics : the journal of
  the american society for experimental neurotherapeutics.
\newblock {\em Journal of the American Society for Experimental
  NeuroTherapeutics}, 9(4):678--701, 2012.

\bibitem{1}
Jens~B Stephansen, Alexander~N Olesen, Mads Olsen, Aditya Ambati, Eileen~B
  Leary, Hyatt~E Moore, Oscar Carrillo, Ling Lin, Fang Han, Han Yan, et~al.
\newblock Neural network analysis of sleep stages enables efficient diagnosis
  of narcolepsy.
\newblock {\em Nature communications}, 9(1):1--15, 2018.

\bibitem{11}
C.~Halasz and Preeyanka Shah.
\newblock Better diagnosing narcolepsy.
\newblock 2013.

\bibitem{28}
Hui~Wen Loh, Chui~Ping Ooi, Jahmunah Vicnesh, Shu~Lih Oh, Oliver Faust,
  Arkadiusz Gertych, and U~Rajendra Acharya.
\newblock Automated detection of sleep stages using deep learning techniques: A
  systematic review of the last decade (2010--2020).
\newblock {\em Applied Sciences}, 10(24):8963, 2020.

\bibitem{29}
F{\'a}bio Mendon{\c{c}}a, Ana Fred, Sheikh~Shanawaz Mostafa, Fernando
  Morgado-Dias, and Antonio~G Ravelo-Garc{\'\i}a.
\newblock Automatic detection of cyclic alternating pattern.
\newblock {\em Neural Computing and Applications}, pages 1--11, 2018.

\bibitem{A1}
Alex Krizhevsky, Ilya Sutskever, and Geoffrey~E Hinton.
\newblock Imagenet classification with deep convolutional neural networks.
\newblock In {\em NIPS'12: Proceedings of the 25th International Conference on
  Neural Information Processing Systems}, page 1097–1105, 2012.

\bibitem{15}
Tianqi Zhu, Wei Luo, and Feng Yu.
\newblock Convolution-and attention-based neural network for automated sleep
  stage classification.
\newblock {\em International Journal of Environmental Research and Public
  Health}, 17(11):4152, 2020.

\bibitem{16}
Zhihong Cui, Xiangwei Zheng, Xuexiao Shao, and Lizhen Cui.
\newblock Automatic sleep stage classification based on convolutional neural
  network and fine-grained segments.
\newblock {\em Complexity}, 2018, 2018.

\bibitem{17}
Ozal Yildirim, Ulas~Baran Baloglu, and U~Rajendra Acharya.
\newblock A deep learning model for automated sleep stages classification using
  psg signals.
\newblock {\em International journal of environmental research and public
  health}, 16(4):599, 2019.

\bibitem{urtnasan2021ai}
E.~Urtnasan, E.Y. Joo, and K.H. Lee.
\newblock Ai-enabled algorithm for automatic classification of sleep disorders
  based on single-lead electrocardiogram.
\newblock {\em Diagnostics}, 11, 2021.

\bibitem{A2}
Urtnasan E., Park J.U., and Lee K.J.
\newblock Multiclass classification of obstructive sleep apnea/hypopnea based
  on a convolutional neural network from a single-lead electrocardiogram.
\newblock {\em Physiol Meas.}, 39(6), 2018.

\bibitem{A3}
Sharma M., Dhiman H.S., and Acharya U.R.
\newblock Automatic identification of insomnia using optimal antisymmetric
  biorthogonal wavelet filter bank with ecg signals.
\newblock {\em Comput Biol Med.}, 131, 2021.

\bibitem{A4}
Han C., Song Y., and Lim H.S.and Tae Y.and Jang J.H.and Lee B.T.and Lee Y.and
  Bae W.and~Yoon D.
\newblock Automated detection of acute myocardial infarction using asynchronous
  electrocardiogram signals-preview of implementing artificial intelligence
  with multichannel electrocardiographs obtained from smartwatches:
  Retrospective study.
\newblock {\em J Med Internet Res.}, 23, 2021.

\bibitem{exarchos2020supervised}
Ioannis Exarchos, Anna~A Rogers, Lauren~M Aiani, Robert~E Gross, Gari~D
  Clifford, Nigel~P Pedersen, and Jon~T Willie.
\newblock Supervised and unsupervised machine learning for automated scoring of
  sleep--wake and cataplexy in a mouse model of narcolepsy.
\newblock {\em Sleep}, 43(5):1--12, 2020.

\bibitem{18}
Huy Phan, Fernando Andreotti, Navin Cooray, Oliver~Y Ch{\'e}n, and Maarten
  De~Vos.
\newblock Joint classification and prediction cnn framework for automatic sleep
  stage classification.
\newblock {\em IEEE Transactions on Biomedical Engineering}, 66(5):1285--1296,
  2018.

\bibitem{19}
Ameni Trabelsi, Mohamed Chaabane, and Asa Ben-Hur.
\newblock Comprehensive evaluation of deep learning architectures for
  prediction of dna/rna sequence binding specificities.
\newblock {\em Bioinformatics}, 35(14):i269--i277, 2019.

\bibitem{20}
Haixia Long, Bo~Liao, Xingyu Xu, and Jialiang Yang.
\newblock A hybrid deep learning model for predicting protein hydroxylation
  sites.
\newblock {\em International journal of molecular sciences}, 19(9):2817, 2018.

\bibitem{14}
Hui~Wen Loh, Chui~Ping Ooi, Jahmunah Vicnesh, Shu~Lih Oh, Oliver Faust,
  Arkadiusz Gertych, and U~Rajendra Acharya.
\newblock Automated detection of sleep stages using deep learning techniques: A
  systematic review of the last decade (2010--2020).
\newblock {\em Applied Sciences}, 10(24):8963, 2020.

\bibitem{24}
Corinna Cortes and Vladimir Vapnik.
\newblock Support vector machine.
\newblock {\em Machine learning}, 20(3):273--297, 1995.

\bibitem{30}
Tin~Kam Ho.
\newblock Random decision forests.
\newblock 1:278--282 vol.1, 1995.

\bibitem{gu2018recent}
Jiuxiang Gu, Zhenhua Wang, Jason Kuen, Lianyang Ma, Amir Shahroudy, Bing Shuai,
  Ting Liu, Xingxing Wang, Gang Wang, Jianfei Cai, et~al.
\newblock Recent advances in convolutional neural networks.
\newblock {\em Pattern Recognition}, 77:354--377, 2018.

\bibitem{31}
Sepp Hochreiter, Yoshua Bengio, Paolo Frasconi, J{\"u}rgen Schmidhuber, et~al.
\newblock Gradient flow in recurrent nets: the difficulty of learning long-term
  dependencies, 2001.

\bibitem{27}
Ramprasaath~R. Selvaraju, Michael Cogswell, Abhishek Das, Ramakrishna Vedantam,
  Devi Parikh, and Dhruv Batra.
\newblock Grad-cam: Visual explanations from deep networks via gradient-based
  localization.
\newblock {\em International Journal of Computer Vision}, 128(2):336–359, Oct
  2019.

\bibitem{ribeiro2016should}
Marco~Tulio Ribeiro, Sameer Singh, and Carlos Guestrin.
\newblock " why should i trust you?" explaining the predictions of any
  classifier.
\newblock In {\em Proceedings of the 22nd ACM SIGKDD international conference
  on knowledge discovery and data mining}, pages 1135--1144, 2016.

\bibitem{25}
Ary~L Goldberger, Luis~AN Amaral, Leon Glass, Jeffrey~M Hausdorff, Plamen~Ch
  Ivanov, Roger~G Mark, Joseph~E Mietus, George~B Moody, Chung-Kang Peng, and
  H~Eugene Stanley.
\newblock Physiobank, physiotoolkit, and physionet: components of a new
  research resource for complex physiologic signals.
\newblock {\em circulation}, 101(23):e215--e220, 2000.

\bibitem{FDA}
Leming Shi, Gregory Campbell, Wendell~D Jones, Fabien Campagne, Zhining Wen,
  Stephen~J Walker, Zhenqiang Su, Tzu-Ming Chu, Federico~M Goodsaid, Lajos
  Pusztai, et~al.
\newblock The microarray quality control (maqc)-ii study of common practices
  for the development and validation of microarray-based predictive models.
\newblock {\em Nature biotechnology}, 28(8):827--838, 2010.

\bibitem{A6}
D~Zhuang and A.K. Ibrahim.
\newblock Deep learning for drug discovery: A study of identifying high
  efficacy drug compounds using a cascade transfer learning approach.
\newblock {\em Applied Sciences}, 11(17), 2021.

\bibitem{RF}
M.and Gunn~S. Saunders, C.and~Grobelnik, Shawe-Taylor, and J.~(eds)~in
  Subspace.
\newblock {\em The {\TeX} Identifying Feature Relevance Using a Random Forest.}
\newblock Springer, Berlin, Heidelberg, 2005.

\bibitem{sainath2015learning}
Tara Sainath, Ron~J Weiss, Kevin Wilson, Andrew~W Senior, and Oriol Vinyals.
\newblock Learning the speech front-end with raw waveform cldnns.
\newblock 2015.

\bibitem{dai2017very}
Wei Dai, Chia Dai, Shuhui Qu, Juncheng Li, and Samarjit Das.
\newblock Very deep convolutional neural networks for raw waveforms.
\newblock In {\em 2017 IEEE international conference on acoustics, speech and
  signal processing (ICASSP)}, pages 421--425. IEEE, 2017.

\bibitem{SALARI2022115950}
Nader Salari, Amin Hosseinian-Far, Masoud Mohammadi, Hooman Ghasemi, Habibolah
  Khazaie, Alireza Daneshkhah, and Arash Ahmadi.
\newblock Detection of sleep apnea using machine learning algorithms based on
  ecg signals: A comprehensive systematic review.
\newblock {\em Expert Systems with Applications}, 187:115950, 2022.

\bibitem{bandyopadhyay2022clinical}
Anuja Bandyopadhyay and Cathy Goldstein.
\newblock Clinical applications of artificial intelligence in sleep medicine: a
  sleep clinician’s perspective.
\newblock {\em Sleep and Breathing}, pages 1--17, 2022.

\end{thebibliography}

\end{document}